\documentclass[onecolumn]{IEEEtran}
\usepackage{xcolor}
\usepackage{cite}
\usepackage{comment}
\usepackage{cite}
\usepackage{graphicx}
\usepackage{amsmath}
\usepackage{amssymb}

\def\XXint#1#2#3{{\setbox0=\hbox{$#1{#2#3}{\int}$}
     \vcenter{\hbox{$#2#3$}}\kern-.5\wd0}}

\newcommand{\be}[1]{\begin{equation} \label{#1} }
\newcommand{\bea}[1]{\begin{eqnarray} \label{#1} }

\newcommand{\bfi}{\begin{figure}}
\newcommand{\efi}{\end{figure}} 
\newcommand{\ee}{\end{equation}}
\newcommand{\eea}{\end{eqnarray}}
\newcommand{\lbl}{\label}
\newcommand{\bib}{\bibitem}

\newcommand{\vl}{{\bf l}}

\newcommand{\vv}{{\bf v}}
\newcommand{\vf}{{\bf f}}

\newcommand{\vzh}{{\bf \hat{z}}}
\newcommand{\vyh}{{\bf \hat{y}}}
\newcommand{\vxh}{{\bf \hat{x}}}

\newcommand{\vr}{{\bf r}}
\newcommand{\vH}{{\bf H}}
\newcommand{\vE}{{\bf E}}

\newcommand{\vP}{{\bf P}}

\newcommand{\vA}{{\bf A}}
\newcommand{\vB}{{\bf B}}
\newcommand{\vF}{{\bf F}}
\newcommand{\vD}{{\bf D}}
\newcommand{\vJ}{{\bf J}}

\newcommand{\vnh}{{\bf \hat{n}}}

\newcommand{\vrho}{{\mbox{\boldmath $\rho$}}}

\newcommand{\eps}{\epsilon}
\begin{document}
\title{Maxwell's derivation of the Lorentz force from Faraday's law}
\author{\IEEEauthorblockN{Arthur D. Yaghjian}}
%\IEEEauthorblockA{Electromagnetics Research Consultant, Concord  MA, USA, a.yaghjian@comcast.net}}
%
\maketitle
\begin{abstract}
In a brief but brilliant derivation that can be found in Maxwell's \textit{Treatise} and traced back to his 1861 and 1865 papers, he derives the force on a moving electric charge subject to electromagnetic fields from his mathematical expression of Faraday's law for a moving circuit.  Maxwell's derivation in his \textit{Treatise} of this force, which is usually referred to today as the Lorentz force, is given in detail in the present paper using Maxwell's same procedure but with more modern notation.
\end{abstract}
%
%\vskip0.5\baselineskip
%\begin{IEEEkeywords}
% electromagnetics, Lorentz force, Maxwell, Faraday's law.
%\end{IEEEkeywords}
%
\section{Introduction}
In Article 603 of his \textit{Treatise} \cite{Maxwell}, \cite{Yaghjian-Reflection}, Maxwell derives the force density $\vf_{\rm c}$ on a conductor carrying electric current density $\vJ$ through a magnetic field $\vB$, namely
\be{1}
\vf_{\rm \,c} = \vJ\times\vB.
\ee
His derivation is based on the mutual induction between two current carrying circuits that represent magnetic shells and nowhere in his \textit{Treatise} does he express $\vJ$ as $\rho\vv$, where $\rho$ is the electric charge density and $\vv$ is the velocity of the charge density. Therefore,  the credit for the force $q\vv\times\vB$ on an electric charge $q$ moving in a magnetic field is generally given to Heaviside \cite{Heaviside1889}  and credit for the total force $q(\vE +\vv\times\vB)$ on a moving charge in electric and magnetic fields is generally given to Lorentz \cite{Lorentz1892}, \cite[app. 7]{Buchwald}.
\par
Nevertheless, it was Maxwell who first determined the general equation for the force on a moving unit electric charge, namely (in our present-day notation)
\be{2}
\vF_{\!\rm unit} = \vE +\vv\times\vB
\ee
in a remarkable derivation from the general equation for Faraday's law that he deduced from Faraday's experiments \cite{Faraday}.\footnote{Faraday did not write any equations in his \textit{Experimental Researches} \cite{Faraday}.  The clearest concise statement that I could find in Faraday's writings on electromagnetic induction (Faraday's law) is in Paragraph 3087 of his \textit{Experimental Researches} \cite{Faraday}, namely, ``The first practical result produced by the apparatus described, in respect of magneto-electric induction generally, is, that a piece of metal or conducting matter which moves across lines of magnetic force, has, or tends to have, a current of electricity produced in it.''  Following this statement, Faraday continues with a more detailed explanation of the ``full effect'' of the experimentally observed magneto-electric induction.}  It is the main purpose of this paper to document and reproduce in modern notation this impressive derivation of Maxwell contained in his \textit{Treatise} \cite{Maxwell}, \cite{Yaghjian-Reflection}\footnote{A shortened version of the derivation in the present paper is given in \cite{Yaghjian-Reflection} but it contains an error pointed out by Red\v{z}i\'{c} \cite{Redzic}.  Red\v{z}i\'{c}'s treatment  \cite{Redzic} of Maxwell's derivation in his \textit{Treatise} of the force on a moving electric charge differs from Maxwell's derivation (and the derivation given here) in that it requires differentiation of the differential of the position vector as well as a separate mathematical proof that the time derivative can be brought inside the line integral of the vector potential for a moving curve.} and, in more rudimentary forms, in his 1861 and 1865 papers \cite{Maxwell1861}, \cite{Maxwell1865}.  Incidentally, in his \textit{Treatise}, Maxwell abandoned the mechanical models he used to develop his equations {\color{black}(mainly in his 1861 paper),} stating that there are an ``infinite number'' of possible ``demonstrations that a mechanism may be imagined capable of producing a connection of the parts of the electromagnetic field'' \cite[art. 831]{Maxwell}.
\section{Maxwell's Mathematical Formulation of Faraday's Law}\label{FL}
In Articles 530--541 of his \textit{Treatise} \cite{Maxwell}, Maxwell explains some of Faraday's experiments by means of ``primary" and ``secondary" circuits that allow him to summarize in Article 541 the ``true law of magneto-electric induction [Faraday's law of induced electromotive force]" as follows:  ``The total electromotive force acting round a circuit at any instant is measured by the rate of decrease of the number of lines of magnetic force which pass through it."  In Chapter IV of Part IV of the \textit{Treatise}, he explains that Faraday's experiments with a single solenoidal circuit also demonstrate a self-induced electromotive force.  %In Chapters V and VI of Part IV, he introduces the Lagrange method that he subsequently uses in Chapters VII--IX of Part IV to deduce a single dynamical theory of electricity that includes self-induced and mutually induced electromotive forces.
\par
Maxwell begins the formulation of time-varying electromagnetic-field equations per se with Chapter VII, ``Theory of Electric Circuits," in Part IV of the \textit{Treatise}.  In this chapter as well as the following Chapter VIII, specifically in Articles 578--592, he culminates a lengthy argument based on the experimental results of Amp\`ere and Faraday with a mathematical formulation of these results in a form we recognize today as Maxwell's first equation.  It is most noteworthy that, although Maxwell does not include his equation for Faraday's law explicitly in his summary of equations in Article 619 because, evidently, he decided finally to emphasize the vector and scalar potential representations of his equations,\footnote{Physicists often note that quantum field theory is indebted to Maxwell's emphasis on the vector and scalar potentials in his final summary of equations in Article 619.} he first wrote down the integral form of Faraday's law in Articles 579 and 595 as %(there should be a minus sign before $Y'$ and $dp/dt$ in the second and third equations of Art. 579)
\be{35}
E(t) = -\frac{d}{dt}p(t)
\ee
where $E(t)$ is the line integral of the dynamic electromotive force per unit electric charge in a circuit (closed curve) that can be moving (and deforming).  For a stationary circuit, $E$ has been given in Article 69 as $\int_C \vE\cdot d\vl$, where $\vE$ is the static ``electric intensity, or electromotive force''.  For a moving circuit, $E$ can be written in terms of a vector $\vE_{\rm v}$\footnote{Maxwell uses the same symbol $\vE$ for the force on a stationary or moving unit electric charge.  Here, as in \cite{Yaghjian-Reflection}, the symbol $\vE$ is reserved for the force on a stationary unit electric charge (the usual electric field) and the symbol  $\vE_{\rm v}$ is used for the force on a unit electric charge moving with velocity $\vv$.  (Of course, $\vE_{\vv=0}=\vE$.)  Also, it should be noted that J. J. Thomson evidently changed the term ``electromotive force'' used by Maxwell in Article 618 and elsewhere in volume two of the first and second editions of Maxwell's \textit{Treatise} to ``electromotive intensity'' in the third edition of Maxwell's \textit{Treatise}.  Whenever this occurs in the quotes of Maxwell used in this paper, I have reinserted Maxwell's original word ``force''.} as \cite[art. 598, eq. (6)]{Maxwell}, \cite[sec. 6.1]{Yaghjian-Reflection}
\be{36}
E(t) = \oint\limits_{C(t)} \vE_{\rm v}(\vr,t) \cdot d\vl 
\ee
where $C(t)$ denotes the closed curve of the moving circuit and $\vE_{\rm v}(\vr,t)$ is the \textit{unknown} unit-charge force that would be exerted on a hypothetical\footnote{The adjective ``hypothetical'' is used herein to denote the force that would be exerted on a small particle carrying a unit of electric charge if it were placed at a point without disturbing the given sources.  In the words that Maxwell used to define the static electric field with a stationary unit charge, ``The resultant electric intensity [field] at any point is the force which would be exerted on a small body charged with the unit of positive electricity, if it were placed there without disturbing the actual distribution of electricity'' \cite[art. 68]{Maxwell}.} electric charge placed at and moving with the point $\vr$ of the circuit (see Articles 68 and 598).   The vector form of the line integral in (\ref{36}) is given in \cite[art. 598, eq. (6)]{Maxwell}.  (It should be noted that in Article 579, $E$ represents the ``impressed" voltage  produced by a battery in the circuit so that $E-IR$ in Article 579 equals $-\oint_{C(t)} \vE_{\rm v} \cdot d\vl$ and thus in Article 579 Maxwell writes $E-IR= dp/dt$.)   
\par
The $p(t)$ in (\ref{35}) is given in Articles 590 and 591 in terms of the vector potential $\vA(\vr,t)$ and magnetic induction field $\vB(\vr,t)$
\be{37}
p(t) = \oint\limits_{C(t)} \vA(\vr,t) \cdot d\vl \; = \int\limits_{S(t)} \vB(\vr,t)\cdot\vnh\, dS 
\ee
with $S(t)$ any open surface bounded by the circuit (closed curve) $C(t)$ and $\vl$ (Maxwell's $\vrho$) is ``the vector from the origin to a point of the circuit."   The vector $\vnh$ is the usual surface unit normal defined by the right-hand rule with respect to the direction $d\vl$ around the curve $C(t)$.   On the curve $C(t)$, the vectors $\vr$ and $\vl$ are identical ($\vr=\vl$ on $C(t)$).  The vector forms of the line and surface integrals in (\ref{37}) are given in \cite[art. 590, eq. (7)]{Maxwell} and \cite[art. 591, eq. (12)]{Maxwell}, respectively.   With (\ref{36}) and (\ref{37}) inserted into (\ref{35}), we see that Maxwell has obtained the most general integral form of Faraday's law
\be{38}
\oint\limits_{C(t)}\!\vE_{\rm v}(\vr,t) \cdot d\vl = -\frac{d}{dt}\!\!\oint\limits_{C(t)} \!\vA(\vr,t) \cdot d\vl = -\frac{d}{dt}\!\!\int\limits_{S(t)}\!\vB(\vr,t)\cdot\vnh\, dS.
\ee
When he writes (\ref{35}) in Article 595, however, he has not yet shown for moving circuits that the electromotive  force ($\vE_{\rm v}$) on a hypothetical unit charge moving with the circuit at $\vr$ is equal to what now is generally called the Lorentz force $\vE +\vv\times \vB$ per moving unit electric charge, where $\vv$ is the velocity of the moving charge.  (Before Maxwell evaluates $\vE_{\rm v}$ to show that it equals $\vE +\vv\times \vB$, he has inferred that $\vE_{\rm v}$ is the electromotive force on a moving unit electric charge from Faraday's measurements \cite{Faraday} of voltages induced in circuits moving through magnetic fields.)
\par
Maxwell allowed his moving ``circuit'' $C(t)$ in (\ref{38}) to be a purely geometrical closed curve moving through the free-space vacuum/ether or probably even through (not just with) a conductor or polarized material.  For example, in Article 598 he references Article 69, where he defines $\int_C\vE\cdot d\vl$ for the static electric field and explains that this integral is ``the work that would be done by the electric force on a unit of positive electricity [electric charge] carried along the curve from $A$, the beginning, to $P$, the end of the arc.''  Moreover, in Article 586, Maxwell says, ``we are not now considering a \textit{current} the parts of which may, and indeed do, act on one another, but a mere \textit{circuit}, that is, a closed curve along which a current \textit{may} flow, and this is a purely geometrical figure, the parts of which cannot be conceived to have any physical action on each other.''  Similar use of the term ``circuit'' to refer to any geometrical ``closed curve'' is found in Articles 587--591, which lead into Article 598.
\par
For stationary circuits, he confirms toward the end of Article 598 that, as in Article 69%, $E(t) =\oint_{C} \vE(\vr,t) \cdot d\vl$, 
\be{stat}
E(t) =\oint\limits_{C} \vE(\vr,t) \cdot d\vl
\ee
where $\vE(\vr,t)$ is the electric force on a hypothetical unit electric charge placed at $\vr$ as explained in Article 68. %he has already concluded from the experiments of Faraday and Amp\`ere (see Arts. 490, 541, and 592) that the electromotive force is equal to the electric field $\vE$ plus a magnetic force proportional to $\vI\times\vB$, where $\vI = I d\vl/dc$ is the directed current in the stationary circuit. 
Consequently, %since $(\vI\times\vB)\cdot d\vl = 0$, 
Maxwell has obtained the integral form of Faraday's law for stationary circuits, namely
\be{39}
\oint\limits_{C} \vE(\vr,t) \cdot d\vl = -\oint\limits_{C} \frac{\partial}{\partial t}\vA(\vr,t) \cdot d\vl  = -\int\limits_{S} \frac{\partial}{\partial t}\vB(\vr,t)\cdot\vnh\, dS.
\ee
Application of Stokes' theorem to (\ref{39}) yields the differential form of Faraday's law
\be{40}
\nabla\times\vE(\vr,t)  = - \nabla\times\frac{\partial}{\partial t}\vA(\vr,t)   = -\frac{\partial}{\partial t}\vB(\vr,t).
\ee
However, Maxwell does not write this differential form of Faraday's law in his \textit{Treatise} nor in his 1865 paper \cite{Maxwell1865} which contain only the integral form of Faraday's law.\footnote{We know that Maxwell deliberately chose to emphasize the integral form of Faraday's law in his \textit{Treatise} and 1865 paper \cite{Maxwell1865} since he had deduced the differential form of this law from his ``theory of molecular vortices" that he used in his 1861 paper \cite{Maxwell1861} to explain Faraday's experimental results \cite{Bucci}.  In Part 2 of that 1861 paper, which contains no integrals, Maxwell wrote the scalar version of $\nabla\times\vE=-\mu\partial \vH/\partial t \,(=-\partial \vB/\partial t)$ as his equation (54).  Maxwell's mathematical formulation of Faraday's law is not contained in Maxwell's earlier 1856 paper \cite{Maxwell1856} in either the integral or differential form.}   The first equation in (\ref{39}) or (\ref{40}) implies that
\be{40'}
\vE(\vr,t) = -\frac{\partial}{\partial t}\vA(\vr,t)-\nabla\psi_e(\vr,t)
\ee
for stationary circuits, where $\psi_e(\vr,t)$ is a time-dependent as well as a spatially dependent scalar potential function.  In Article 598 Maxwell says that $\psi_e(\vr,t)$ ``represents, according to a certain definition, the \textit{electric potential},"  which he later says in Article 783 satisfies Poisson's equation $\nabla^2 \psi_e(\vr,t) = -\rho(\vr,t)/\eps$ in a homogeneous isotropic dielectric with permittivity $\eps$.  This Poisson equation follows from Maxwell using the Coulomb gauge $\nabla\cdot\vA = 0$ throughout his \textit{Treatise} \cite[secs. 6.2 and 6.4]{Yaghjian-Reflection}.  Maxwell also writes (\ref{40'}) explicitly as his second equation in Article 783 for time varying fields with $\vv=0$.
\section{Derivation of the Force on a Moving Unit Electric Charge}
Returning to Maxwell's general integral form of Faraday's law for moving circuits in (\ref{35}), expressed more fully in (\ref{38}), we find in Article 598 Maxwell's ingenious evaluation of $-(d/dt)\oint_{C(t)} \vA(\vr,t) \cdot d\vl$ to prove that $\vE_{\rm v}(\vr,t)= -\partial\vA(\vr,t)/\partial t-\nabla\psi_e(\vr,t) +\vv(\vr,t)\times \vB(\vr,t)$, where $\vv(\vr,t)$ is the velocity at each point $\vr=\vl$ of the moving circuit $C(t)$.  He thus completes the mathematical formulation of Faraday's integral law for moving circuits and in so doing derives the force exerted on a moving unit electric charge by the magnetic induction field $\vB$.  Maxwell accomplishes this feat as follows.
\par
He writes $\vA(\vr,t) \cdot d\vl$ in rectangular coordinates as
\be{41}
\vA(\vr,t) \cdot d\vl = A_x \frac{\partial x}{\partial s}ds +A_y \frac{\partial y}{\partial s}ds+A_z \frac{\partial z}{\partial s}ds 
\ee
where $s$ is defined by Maxwell in Articles 16 and 69 as ``the length of the arc, measured from $A$ [the initial point on the arc]''.  Thus, $ds$ is the scalar element of length on the closed curve $C(t)$ at a fixed time $t$.  The $x$, $y$, and $z$ are the rectangular components of the position vector $\vl = \vr = x\vxh+y\vyh+z\vzh$ on $C(t)$.  These rectangular components are functions of the time $t$ and the length $s$ along the curve $C(t)$, that is
\be{rxyz}
\vl(t,s) = \vr(t,s) = x(t,s)\vxh+y(t,s)\vyh+z(t,s)\vzh,\;\;\;\; \vl=\vr \in C(t).
\ee
\par
If we consider the integral
\be{41'}
\oint\limits_{C(t)}\vA(\vr,t) \cdot d\vl = \oint\limits_{C(t)}\left(A_x \frac{\partial x}{\partial s} +A_y \frac{\partial y}{\partial s} +A_z\frac{\partial z}{\partial s}\right)ds
\ee
one can change the scalar integration variable $s$ to $s' = s/s_{max}$ at each instant of time $t$, where $s_{max}$ is the total length of the closed curve $C(t)$ at the time $t$, and (\ref{41'}) becomes
\be{41''}
\oint\limits_{C(t)}\vA(\vr,t) \cdot d\vl = \oint\limits_0^1\left(A_x \frac{\partial x}{\partial s'} +A_y \frac{\partial y}{\partial s'} +A_z \frac{\partial z}{\partial s'}\right)ds'
\ee
and thus the limits of the integration variable $s'$ need not change with time $t$.
Since $C(t)$ is a closed curve, $s'=0$ and $s'=1$ refer to the same point on $C(t)$.  Maxwell didn't do this renormalization of the $s$ variable explicitly because it was probably obvious to him that the limits of the integration variable $s$ can be chosen to be independent of the time variable $t$ since it occurs in both the numerator and denominator of the right hand side of (\ref{41}).
\par
Taking the time derivative of this equation, we can bring the time derivative of the right-hand side under the integral sign (because the limits of integration do not depend on time) to get
\bea{42}
&&\hspace{-5mm}\frac{d}{dt}\oint\limits_{C(t)} \vA(\vr,t) \cdot d\vl =\oint\limits_0^1 \frac{d}{dt}\left[A_x\big[\vr(t,s),t\big] \frac{\partial x}{\partial s} + \cdots %A_y\big[\vr(t,c),t\big] \frac{\partial y}{\partial s} + A_z\big[\vr(t,c),t\big] \frac{\partial z}{\partial s}
\right]ds\nonumber\\
&&=\oint\limits_0^1 \bigg[\frac{\partial A_x(\vr,t)}{\partial t}  \frac{\partial x}{\partial s}+\frac{\partial A_y(\vr,t)}{\partial t}\frac{\partial y }{\partial s}+\frac{\partial A_z(\vr,t)}{\partial t}  \frac{\partial z}{\partial s}\nonumber\\
&&+\left(\frac{\partial A_x}{\partial x}\frac{\partial x}{\partial s} + \frac{\partial A_y}{\partial x}\frac{\partial y}{\partial s} + \frac{\partial A_z}{\partial x}\frac{\partial z}{\partial s}\right)\frac{\partial x}{\partial t}\\
&&+\left(\frac{\partial A_x}{\partial y}\frac{\partial x}{\partial s} + \frac{\partial A_y}{\partial y}\frac{\partial y}{\partial s} + \frac{\partial A_z}{\partial y}\frac{\partial z}{\partial s}\right)\frac{\partial y}{\partial t}\nonumber\\
&&+\left(\frac{\partial A_x}{\partial z}\frac{\partial x}{\partial s} + \frac{\partial A_y}{\partial z}\frac{\partial y}{\partial s} + \frac{\partial A_z}{\partial z}\frac{\partial z}{\partial s}\right)\frac{\partial z}{\partial t}\nonumber\\
&&+ \left(A_x(\vr,t)\frac{\partial^2 x}{\partial s \partial t} + A_y(\vr,t)\frac{\partial^2 y}{\partial s \partial t} + A_z(\vr,t)\frac{\partial^2 z}{\partial s \partial t}\right)\bigg] ds\nonumber
\eea
where the superfluous prime on the integration variable $s'$ has been dropped.
The partial derivatives with respect to $s$ are taken holding $t$ fixed.  The partial derivatives of $(x,y,z)$ with respect to $t$ are taken holding $s$ fixed.  The partial derivatives of $\vA(\vr,t)$ with respect to $x$, $y$, or $z$ are taken holding $t$ fixed and the partial $t$ derivative of $\vA(\vr,t)$ is taken holding $(x,y,z)$ fixed.
To obtain (\ref{42}), use has been made of the chain rules
\be{cr1}
\frac{d\vA(\vr,t)}{dt}=\frac{\partial\vA(\vr,t)}{\partial t}+\frac{\partial\vA(\vr,t)}{\partial x}\frac{dx}{dt}
+\frac{\partial\vA}{\partial y}\frac{dy}{dt}+\frac{\partial\vA}{\partial z}\frac{dz}{dt}
\ee
\begin{subequations}
\lbl{cr2}
\be{cr2a}
\frac{dx(t,s)}{dt}=\frac{\partial x(t,s)}{\partial t}+\frac{\partial x(t,s)}{\partial s}\frac{ds}{dt} = \frac{\partial x(t,s)}{\partial t}=v_x
\ee
\mbox{}\\[-6mm]
\be{cr2b}
\frac{dy(t,s)}{dt}=\frac{\partial y(t,s)}{\partial t} = v_y
\ee
\be{cr2c}
\frac{dz(t,s)}{dt}=\frac{\partial z(t,s)}{\partial t} = v_z
\ee
\end{subequations}
while noting that the variable $s$ is independent of time ($ds/dt=0$) since it varies from $0$ to $1$ independently of $t$.  The chain rule in (\ref{cr1}) holds for any $dx$, $dy$, and $dz$ so we can choose $(x,y,z)$ to be the coordinates $[x(t,s),y(t,s),z(t,s)]$ of the curve whose length has been normalized to $1$.
\par
If we proceed as Maxwell did, using $\vB = \nabla\times\vA$ to substitute $\partial A_y/\partial x = \partial A_x/\partial y + B_z$ and $\partial A_z/\partial x = \partial A_x/\partial z - B_y$ into the third line of (\ref{42}), we get for that line
\bea{M239-1}
&&\hspace{-6mm}\oint\limits_0^1\left(B_z\frac{\partial y}{\partial s} - B_y\frac{\partial z}{\partial s} + \frac{\partial A_x}{\partial x}\frac{\partial x}{\partial s} + \frac{\partial A_x}{\partial y}\frac{\partial y}{\partial s} + \frac{\partial A_x}{\partial z}\frac{\partial z}{\partial s}\right)\frac{\partial x}{\partial t} ds\nonumber\\[-2mm]
&&\hspace{15mm}= \oint\limits_0^1\left(B_z\frac{\partial y}{\partial s} - B_y\frac{\partial z}{\partial s} + \frac{\partial A_x}{\partial s}\right)\frac{\partial x}{\partial t} ds.
\eea
Because
\be{id2}
\frac{\partial A_x}{\partial s}\frac{\partial x}{\partial t}+ A_x\frac{\partial^2 x}{\partial s \partial t}=\frac{\partial}{\partial s}\left(A_x\frac{\partial x}{\partial t}\right)
\ee
is a perfect differential, its integral around the closed curve of unity length is zero.  Thus, (\ref{M239-1}), along with the similar expressions for the fourth and fifth lines in (\ref{42}), reduce (\ref{42}) to
\bea{43}
&&\hspace{-6mm}\frac{d}{dt}\!\oint\limits_{C(t)}\! \vA(\vr,t) \cdot d\vl =\oint\limits_0^1\left[\left(\frac{\partial A_x(\vr,t)}{\partial t}  + B_y\frac{\partial z}{\partial t}-B_z\frac{\partial y}{\partial t} \right)\frac{\partial x}{\partial s}\right.\nonumber\\
&&\hspace{10mm}+\left(\frac{\partial A_y(\vr,t)}{\partial t}  + B_z\frac{\partial x}{\partial t}-B_x\frac{\partial z}{\partial t} \right)\frac{\partial y}{\partial s}\nonumber\\
&&\hspace{10mm}\left.+\left(\frac{\partial A_z(\vr,t)}{\partial t}  + B_x\frac{\partial y}{\partial t}-B_y\frac{\partial x}{\partial t} \right)\frac{\partial z}{\partial s}\right]ds\nonumber\\[3mm]
&&\hspace{10mm}=\oint\limits_{C(t)}\bigg[\frac{\partial}{\partial t}\vA(\vr,t)-\vv(\vr,t)\times \vB(\vr,t)\bigg]\cdot d\vl 
\eea
where $\vv = \partial x/\partial t\, \vxh + \partial y/\partial t \,\vyh + \partial z/\partial t \,\vzh$.
Consequently, Maxwell has proven that
\be{44}
\oint\limits_{C(t)}\vE_{\rm v}(\vr,t)\cdot d\vl = -\oint\limits_{C(t)}\left[\frac{\partial}{\partial t}\vA(\vr,t) -\vv(\vr,t)\times\vB(\vr,t)\right]\cdot d\vl
\ee
and, thus, he concludes that
\be{450}
\vE_{\rm v}(\vr,t) = -\frac{\partial}{\partial t}\vA(\vr,t)-\nabla\psi_e(\vr,t) +\vv(\vr,t)\times\vB(\vr,t)
\ee
or in accordance with (\ref{40'}) (and our present-day notation $\vE$ for the electric field)
\be{45}
\vE_{\rm v}(\vr,t) = \vE(\vr,t) +\vv(\vr,t)\times\vB(\vr,t)
\ee
since $\vE_{\rm v}$ in (\ref{450}) with $\vv=0$ has to equal $\vE$ in (\ref{40'}).  As mentioned above, Maxwell writes $\vE = -\partial\vA/\partial t-\nabla\psi_e$ explicitly in Article 783 for time varying fields with $\vv=0$.   It should be emphasized that Maxwell uses his mathematical formulation of Faraday's law to obtain (\ref{450}) and not the conductor current force density $\vJ\times\vB$ in (\ref{1}) that he has found from the force on a magnetic-shell model of circulating electric current.
\par
Therefore, Maxwell  has been able to represent Faraday's experimental results in a general mathematical form of Faraday's law given in (\ref{38}) with $\vE_{\rm v}$ given in (\ref{450}) \cite[arts. 598--599]{Maxwell}. \textit{In one magnificent synthesis of mathematical and physical insight, he has not only put Faraday's law on a solid mathematical foundation but he has also derived the ``Lorentz force'' for a moving unit electric charge.} %(As explained in \cite[sec. 3]{Yaghjian-Reflection},  Maxwell uses the same boldface German letter $\mathfrak{E}$ for both the symbols $\vE_{\rm v}$ and $\vE$ that we use here, and when Maxwell denotes $\vE_{\rm v}$ by $\mathfrak{E}$ he expects the reader to know from the context that $\mathfrak{E}$ with $\vv=0$ is the electric field defined in Articles 44 and 68 and given here as $\vE$ in equation (\ref{40'}).) 
\par 
It is also possible to prove (\ref{38}) and (\ref{450})--(\ref{45}) from (\ref{39}) using the Helmholtz transport theorem \cite[ch. 6]{Tai} of vector calculus, but Maxwell does not do this even though he mentions Helmholtz's work with moving circuits in Article 544.  Effectively, he proves the Helmholtz transport theorem for the electromagnetic fields in his mathematical formulation of Faraday's law as part of his derivation reproduced above in (\ref{41'})--(\ref{450}).
\par
After deriving  (\ref{450}) from (\ref{38}) in Article 598, he says that $\vE_{\rm v}(\vr,t)$ in (\ref{450}) is the most general form of the electromotive force on a hypothetical unit point electric charge moving with $C(t)$ at $\vr$, ``being the force which would be experienced by a [moving] unit of positive electricity [electric charge] at that point.''   It follows from linear superposition that the force on an electric charge $q$ moving through electromagnetic fields is given by (in our present-day notation)
\be{Lf}
\vF = q(\vE + \vv\times\vB)
\ee
what we refer to today as the Lorentz force.
\par
Red\v{z}i\'c \cite{Redzic} suggests that because ``Maxwell's $\psi_e(\vr,t)$ does not have the same connotation as today's scalar potential'' that satisfies the Lorenz-Lorentz gauge, Maxwell may not have interpreted $-\partial \vA(\vr,t)/\partial t - \nabla\psi_e(\vr,t)$ as today's electric field vector $\vE(\vr,t)$ being the force that would be exerted on a hypothetical stationary unit electric charge placed at the point $\vr$.  This is highly unlikely given that Maxwell defines the static electric field $\vE(\vr)$ in Article 68 as the force that would be exerted on a hypothetical stationary unit electric charge at the point $\vr$ and that Maxwell refers to Article 68 in his Article 598.  Moreover, Maxwell always used the Coulomb gauge $\nabla\cdot\vA = 0$ \cite[secs. 6.2 and 6.4]{Yaghjian-Reflection} and, thus, his scalar potential always satisfies $\nabla^2\psi_e(\vr,t)=-\rho(\vr,t)/\eps$ \cite[art. 783]{Maxwell} but his electric force on a stationary unit electric charge is still given by $\vE(\vr,t) =-\partial \vA(\vr,t)/\partial t - \nabla\psi_e(\vr,t)$, a relationship that holds independently of the gauge and that Maxwell writes down explicitly for time varying fields in Article 783.  Although Maxwell explained in Article 599 that the vector called $\vE_{\rm v}(\vr,t)$ herein is also the force experienced by the electric-polarization and conduction charges of a material body (which could be the ether) moving with the curve $C(t)$ as confirmed by his writing in Articles 608, 609, and 619 that $\vD = \eps\vE_{\rm v}$ (Maxwell looked at $\vD$ as electric polarization \cite{Yaghjian-Polarization}) and $\vJ = \sigma\vE_{\rm v}$ (correct for $v^2/c^2 \ll 1$), it seems clear from what Maxwell wrote in Articles 598 and 599 that he fully realized that $-\partial \vA(\vr,t)/\partial t - \nabla\psi_e(\vr,t)$ was the force exerted on a hypothetical stationary unit electric charge placed at the point $\vr$ and that $-\partial \vA(\vr,t)/\partial t - \nabla\psi_e(\vr,t) + \vv\times\vB$ was the force exerted on a hypothetical moving (with velocity $\vv$) unit electric charge placed at the point $\vr$, even though he did not explicitly write $\vE_{\rm v}(\vr,t) = \vE(\vr,t) +\vv(\vr,t)\times\vB(\vr,t)$ as is done in (\ref{45}).
\par
For example, near the end of Article 598, Maxwell says that ``Hence we may now disregard the circumstance that $ds$ forms part of a circuit, and consider it simply as a portion of a moving body, acted on by the electromotive force [my $\vE_{\rm v}$].  The electromotive force has already been defined in Art. 68.  It is also called the resultant electrical force, being the force which would be experienced by a unit of positive electricity [electric charge] placed at that point.  We have now obtained the most general value of this quantity in the case of a body moving in a magnetic field due to a variable electric system.''  He continues in Article 599 with, ``The electromotive force  [on a particle with unit electric charge], the components of which are defined by equations (B) [(\ref{450}) above], depends on three circumstances.  The first of these is the motion of the particle [carrying unit electric charge] through the magnetic field [$\vB$].  The part of the force  depending on this motion is expressed by the first two terms on the right of each equation [$\vv\times\vB$ in (\ref{450})].''  
\par
From these and other statements in his \textit{Treatise}, it seems clear that Maxwell realized that he had derived the force exerted by the electromagnetic fields in free space (ether) {\color{black}or} conductors or polarized material\footnote{Maxwell does not deal with the question of how a hypothetical moving electric charge placed instantaneously at $\vr$ could measure the force $\vE+\vv\times\vB$ for $C(t)$ in a polarized material.  In magnetic polarization (magnetization), he determines his mathematically defined (macroscopic) magnetic fields $[\vB,\vH]$ from his primary free-space magnetic field $\vH_0$ measured in small cavities.  He does not give an analogous  prescription for determining the mathematically defined electric field from cavity fields in polarized dielectrics since Maxwell considered $\vD$ to be the electric polarization and did not introduce a polarization vector $\vP$ \cite{Yaghjian-Reflection}, \cite{Yaghjian-Polarization}.} at each instant of time on a hypothetical  moving particle at $\vr$ carrying unit electric charge\footnote{Even though Maxwell derived the force on electric charge moving through electromagnetic fields, apparently most of the scientific community did not understand what he had done until much later when Lorentz used this force extensively in his work.  In fact, Maxwell's equations were not widely accepted until after Hertz experimentally demonstrated the existence of wireless microwave radiation \cite{Hertz}.  {\color{black}This reluctance to accept Maxwell's equations until well after he had died, Maxwell's direct but mathematically sophisticated deduction of the $\vv\times\vB$ force in Faraday's law in Article 598, and his use of potentials and the same symbol $\vE$ for the force on both a stationary and moving unit electric charge are probably the main reasons that Maxwell's derivation of the ``Lorentz force'' continues to be largely overlooked.}} and, if the velocity $\vv$ of the circuit were zero, the measured time varying force at a point $\vr$ on $C$ would be the time varying electric field $\vE(\vr,t) =-\partial \vA(\vr,t)/\partial t - \nabla\psi_e(\vr,t)$.  Moreover, he explicitly writes this equation for time varying fields and $\vv=0$ in Article 783.   Indeed, if this were not the case, it would mean that for stationary circuits $C$,  Maxwell's vector $\vE(\vr,t)$ in his equation (\ref{38}) of Faraday's law would not refer to the time varying electric field as measured by the force on a hypothetical stationary unit electric charge placed at $\vr$.
\par
Maxwell used the same symbol ``$\vE$'' for the force on a unit electric charge whether or not the charge was moving.  This does not conform to our present-day notation for the electric field but it does not represent a mistake in either his equations (reproduced herein) derived from Faraday's law or his physical interpretation of these equations in terms of the force on moving and stationary unit electric charges.  Other scientists used the same symbol for the force on a moving or stationary unit charge many years after Maxwell's \textit{Treatise}, notably, Poynting in his classic paper deriving what is now known as Poynting's theorem \cite{Poynting}, \cite{Yaghjian-Classical}. 
\section{Conclusion}
From the experiments of Faraday, Maxwell infers that the integral of the electromotive force around a moving circuit (closed curve) $C(t)$ is given by the negative time derivative of the magnetic flux through any open surface $S(t)$ bounded by the closed curve $C(t)$.  Using this generalized mathematical formulation of Faraday's law given in (\ref{38}), Maxwell effectively derives a Helmholtz transport theorem for the electromagnetic fields to prove that the electromagnetic force on a moving electric charge is given by the ``Lorentz force'' in (\ref{Lf}) (using our present-day notation).  This remarkable result derived in Maxwell's \textit{Treatise} can be traced back to his 1861 paper \cite{Maxwell1861}, which was written about 30 years before Heaviside \cite{Heaviside1889} and Lorentz \cite{Lorentz1892} expressed the force on a moving electric charge.
\section*{Acknowledgment}
The manuscript benefited from helpful discussions with Professor Dragan V. Red\v{z}i\'{c} (who does not fully concur in\cite{Redzic} with some of the views presented here).   This work was supported in part under the U.S. Air Force Office of Scientific Research Contract \# FA9550-19-1-0097 through Dr. Arje Nachman.
%
%\linespread{1.0}

%
\end{document}